# The Limits to Machine Consciousness

Subhash Kak
Oklahoma State University

**Abstract** It is generally accepted that machines can replicate cognitive tasks performed by conscious agents as long as they are not based on the capacity of awareness. We consider several views on the nature of subjective awareness, which is fundamental for self-reflection and review, and present reasons why this property is not computable. We argue that consciousness is more than an epiphenomenon and assuming it to be a separate category is consistent with both quantum mechanics and cognitive science. We speak of two kinds of consciousness, little-C and big-C, and discuss the significance of this classification in analyzing the current academic debates in the field. The interaction between the system and the measuring apparatus of the experimenter is examined both from the perspectives of decoherence and the quantum Zeno effect. These ideas are used as context to address the question of limits to machine consciousness.

## 1. INTRODUCTION

AI and robotics are bringing about revolutionary changes in society and the next question is whether machines with consciousness could be designed. The word "consciousness" can mean different things, but in its quest for machines the sense is of "awareness" and "subjectivity" that underlies memories, creating a narrative that goes beyond the straitjacket of physical law [1],[2]. This is also an important scientific question for all cognitions and the creation of science takes place in consciousness. It follows that an understanding of reality cannot emerge without an insight into its nature.

An immediate practical motivation for this research is the prospect that "self-conscious" robots will be deployed on the battlefield and used in rescue operations in dangerous environments. It is clear that machines with awareness will have greater autonomy and corresponding beneficial uses but they will create new challenges by replacing humans in many jobs and raise thorny problems of ethics and morality. Whether such machines will ever be built remains an open question.

A distinction is generally made between the philosophical positions of strong and weak AI. Those who believe in the former admit the possibility that AI will subsume the phenomenon of consciousness, whereas those who believe in the latter allow only the possibility of mimicking the capacities of the brain.

At first look, one could assert that since the brain is a machine that is conscious then other machines with appropriate architecture should also exhibit consciousness. Implicit in this belief is the physicalist position that consciousness is either an emergent property or an epiphenomenon, and that computers can capture the abstract causal organization of other systems and thus of the brain [3],[4]. But this is unfalsifiable for if machines that can do so do not exist now – and they do not – one can always point to the future when such machines will emerge. Panpsychism, the position that mind is to be found everywhere, is another position that has recently become popular in academic circles [5], but it is too extravagant in associating mind even with a straw or a rock. Broadly speaking, both physicalism and panpsychism associate mind with matter, although they do so in different ways.

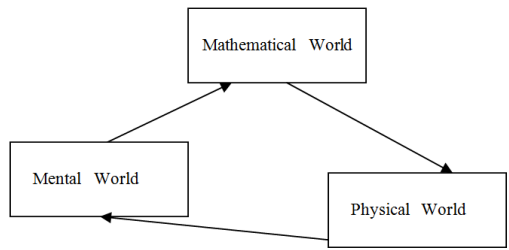

Figure 1. The physical, the mental, and the mathematical worlds

From a computational perspective, it is astonishing that a small subset of the processes in the brain, which we label abstract thought, is able to capture the workings of the physical world to a great degree [6] (Figure 1). It is also noteworthy that brain function is accompanied by the reorganization of its very structures during its learning [7],[8] that goes beyond the commonsense notion of procedure-based computation at the basis of the Turing machine model on which the von Neumann architecture of digital computers is broadly based [9] (Figure 2).

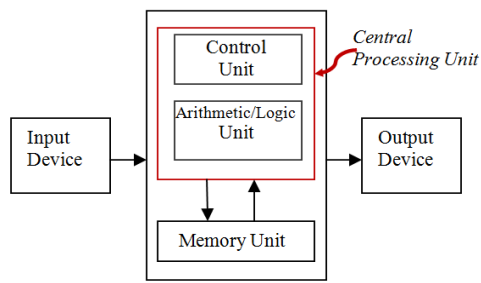

Figure 2. The von Neumann architecture of the computer

Although learning cognitive tasks may require attention and concentration, once the learning is complete they can be performed literally automatically. The learning leads to a training of neural networks that convert the computational problem into one of classification and recognition. It should thus be possible to replicate the tasks that involve computations in the neural circuitry of the brain by means of processing in a machine. But "raw awareness" appears to be a phenomenon that is different from cognitive processing.

This paper begins by summarizing the case for and against a unitary representation of the world and investigates if self-awareness is computable. It is argued that the orthodox Copenhagen Interpretation of quantum mechanics implies that consciousness is a category different from matter which is consistent with the view that awareness is different from the contents of the mind. The interaction between system and the measuring apparatus of the experimenter is examined both from the perspectives of decoherence and the quantum Zeno effect. Lastly, the implications of this analysis for the limitations of machine consciousness are examined.

2. **ONE WORLD OR MANY?**

The motivation for seeing the world in unitary terms where it is reducible to physics derives, in part, from an abhorrence of dualism. Popper and Eccles stressed the shortcomings of the unitary physicalist position in a proposal for interactionism with three worlds as constituents [10],[11]. They saw World 1 as consisting of physical objects and events including biological entities, World 2 as composed of mental objects and events, and World 3 as consisting of objective knowledge. They claimed that World 3 is different from Platonic ideas for it is created by World 2 and it does not have a privileged position.

The three worlds of Popper and Eccles are similar to the old Indian model of Figure 3 that sees the universe in five systems of the body, mediating processes, mind, intuition and science, aesthetics and emotions, together with consciousness as a separate category [12].

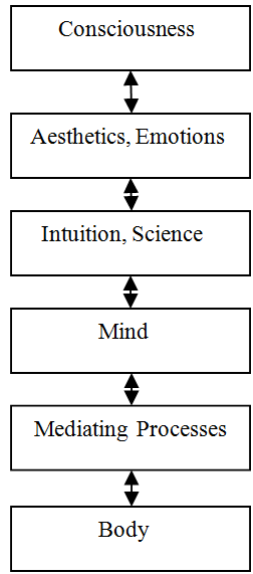

Figure 3. Five planes of existence and consciousness [12]

The traditional Vedic model of the mind and consciousness, which is described at length in the Upanishads and other old texts, is given in Figure 4 and it includes elements such as ego (the sense of the autobiographical self) and awareness. Mind in this model is an inner processor that does operations associated with different cognitive capacities in a systematic manner. Such a model cannot be implemented since we don't know how to implement ego and awareness in a formal system.

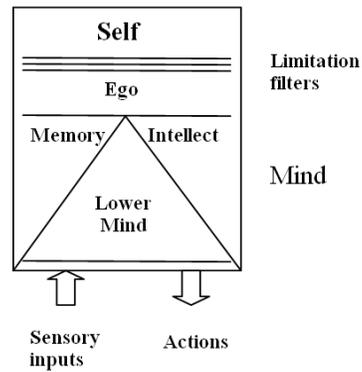

Figure 4. A model of mind and consciousness [12]

The design of a conscious machine faces formidable scientific and engineering obstacles and so one must begin with small steps. Architectures that copy models of brain function have been investigated [2],[13],[14],[15]. These architectures include distributive agents and the global workspace theory (GWT) [16],[17]. In the GWT, separate parallel processes compete to place their information in the global workspace whose contents are broadcast to a multitude of receiving processes. Since globally broadcast messages can evoke actions in receiving processes throughout the network, the global workspace may be used to exercise executive control to perform voluntary actions. This roughly mimics some brain functions without addressing the origin of consciousness.

Postulating a physical substrate of consciousness to be at the basis of complex patterns of activity has been called the integrated information theory (IIT) [18],[19]. But it amounts to correlations in physical processes that cannot by themselves be the source of awareness. Complex causally connected behavioral patterns may also be seen in social networks, where they are properly analyzed within the framework of an ecological system without recourse to the concept of consciousness.

In the standard neuroscience view, mind emerges from the interoperation of the various modules in the brain and its behavior must be completely described by the corresponding brain function (Figure 5). However, no specific neural correlate of consciousness has been found [20]. Others have argued that counterintuitive characteristics of the mind are ascribable to underlying quantum processes [21],[22]. But although quantum mechanics might play a role in brain processes [23],[24] there is no reason to assume that it throws any light on the phenomenon of consciousness [25].

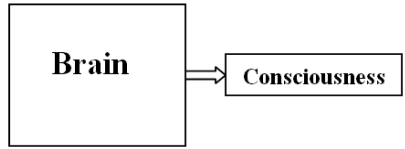

Figure 5. Consciousness as an emergent process

The other position is that consciousness is a fundamental category that is dual to physical reality as in Figure 6 [26],[27]. In art and philosophy, the idea that consciousness is universal is widely accepted. It has been said that the eyes see only what the mind is trained to comprehend and, therefore, culture serves as a filter that structures reality. Technology and inner development have guided the appreciation of modes of artistic expression and the unprecedented changes of the past couple of centuries have led to corresponding efflorescence of art. Just as personal insight is preceded by a crisis, the evolution of artistic consciousness at the societal level comes at the end of a critical phase of doubt. Indeed the themes of *fin de siècle* are seen to prefigure the geopolitical changes of the twentieth century.

We apprehend reality in our consciousness and not directly in terms of space, time and matter. Consciousness is the doorway that shows us the world and makes self-knowledge possible and it is the source of creativity although it is constrained by the habits and limitations of the mind. The quality of the manifestation of consciousness in a natural system depends on structure and different modes of processing.

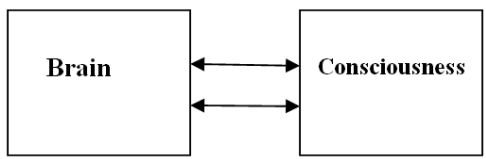

Figure 6. Consciousness as an independent category

Consciousness and the material world complement each other and consciousness may influence material evolution as in the quantum Zeno effect [28],[29]. This is the position of the orthodox Copenhagen Interpretation of quantum mechanics, which is sometimes criticized for supporting dualism. The pioneers of quantum mechanics thought this criticism was misplaced. Noting that Schrödinger believed that the philosophical basis of quantum theory is consistent with the Vedic system (e.g. [30]), we add that in the mainstream Indian philosophical position this dualism is only apparent and at the basis of all is a unitary consciousness [31],[32].

### 3. ON MIND AND COMPUTABILITY

Different observers perceive the same object, say a flower placed on a table, differently based on their perspective and prior state of mind, even though the flower is situated in the same physical space. If one were to switch categories, and imagine consciousness to be like that flower even though it is not a thing, then one can see how it may be variously experienced in different minds. Its apparent plurality is a consequence of its many projections.

The statement that consciousness is the ground on which experience is evoked variously is like saying that of gravity is one force that works on objects differently depending upon their location.

Schrödinger addresses the question of the nature of unity of consciousness in very clear terms [33]:

> Consciousness is never experienced in the plural, only in the singular. Even in the pathological cases of split consciousness or double personality the two persons alternate, they are never manifest simultaneously. ...How does the idea of plurality (so emphatically opposed by the Upanishad writers) arise at all? Consciousness finds itself intimately connected with, and dependent on, the physical state of a limited region of matter, the body. (Consider the changes of mind during the development of the body, as puberty, ageing, dotage, etc., or consider the effects of fever, intoxication, narcosis, lesion of the brain and so on.) Now, there is a great plurality of similar bodies. Hence the pluralization of consciousnesses or minds seems a very suggestive hypothesis. Probably all simple, ingenuous people, as well as the great majority of Western philosophers, have accepted it.

Memories and experiences are not physical although they may have neural correlates. The same awareness of the individual suffering from dissociative identity disorder is bound to different alters (alternate selves). More dramatically, the same individual might be a loving family man at home and a ruthless murderer in a different environment. The self is past experience together with culturally determined ways of binding this experience. Since the wholeness of the self is not diminished if the individual were to be taken to an entirely new society and cut-off from all past experience, the self is beyond the experience.

We think of ourselves as being outside of the physical world. Even our conceptions of the universe are as if we are not a part of it, and in the words of Schrödinger [33]: "We do not belong to this material world that science constructs for us. We are not in it; we are outside. We are only spectators. The reason why we believe that we are in it, that we belong to the picture, is that our bodies are in the picture. Our bodies belong to it."

If this sense of being outside of the physical world is true, it would be impossible to emulate it by hardware and processing that is within the world. It also follows that it will not be a computational property of the physical elements that comprise the system. Penrose thought that consciousness must be non-algorithmic because it is global phenomenon [9],[21].

Can consciousness be seen as the capacity to know with certainty that one is conscious [34],[35]? This appears to be a circular definition and it hinges on hard-to-define concepts such as "knowledge" and "belief". Consider HM, the guy who lost all new memory with the bilateral resection of medial temporal lobe and his ability to hold beliefs and knowledge was greatly impaired [36], yet, without doubt, he had all hallmarks of consciousness.

We can look for the non-computability of consciousness from its parallel to the unsolvability of the halting problem. Let us informally define "consciousness" as some privileged state of the mind that makes its processes halt (we don't bother to specify it beyond this description) and its contents registered (which is what we imply by awareness). Humans can get into this state at any time, which means that the earlier computation has halted, and this is irrespective of the initial state of the immediately preceding process (the exceptions to this are if a person is sleeping or unconscious as in coma). But such halting to arbitrary input is impossible from a computability point of view. Therefore, it follows that consciousness is not computable.

### 4.  LITTLE-C AND BIG-C CONSCIOUSNESS

Two possibilities with regard to the nature of consciousness have been a part of the discourse in many cultures. Nowhere has the debate over these been as sustained and clear as in India where the Vedic tradition represents the view that consciousness is a separate transcendent category (*ātman*) that is apart from physical reality (big-C) and to Buddhist thought is ascribed the view that consciousness is evanescent and fleeting and the underlying reality is emptiness (*śūnyatā*) (little-C). Many scholars believe that the difference between the two positions is more apparent than real (e.g. [37]) for the Buddha speaks of the inability of language to describe the truth (for example, in the Diamond Sutra and the Abhidhamma) indicating its basis is paradoxical which is quite consistent with the Vedic view [38]. According to the Vedic tradition, Buddhist practices are fine in advancing the understanding of the mind, but they stop short by denying the independent agency of consciousness. The idea of impermanence is critiqued on the premise that impermanent reality cannot be constrained by permanent laws and if the laws are permanent they represent a transcendent aspect of the reality.

Formally, big-C is consciousness with an ontological category of its own (e.g. [38]). It appears that consciousness as conceived in psychophysical parallelism is big-C [40], and Schrödinger stressed this point repeatedly (see, e.g. [41]), and the positivist view of reality supports this view.

We use the term little-C for consciousness that relates to the normal functions of the mind that appear as an epiphenomenon as seen, for example, through the lens of neuroscience. The centrality of the non-self doctrine in Buddhist thought, even allowing for the invariant principles that guide mind's transformations [42],[43], is consistent with little-C. These two views are summarized in Table 1.

Table 1. Two views of consciousness

|  | Big-C | Little-C |
|---|---|---|
| Type | Independent category | Epiphenomenon |
| Knowledge | Epistemic | Ontic |
| Systems perspective | Global, integrative | Local, reductionist |
| Logical perspective | Top-down | Bottom-up |

The idea of big-C consciousness raises the question of how it is related to matter and how they mutually influence each other. This is not a problem in little-C consciousness where one may claim that the agency associated with consciousness is an illusion [44].

One might argue that the difference between the two ideas of consciousness is more semantic than real as their self-understanding practices are practically identical. But the two can indeed be distinguished if there are aspects to big-C that cannot be explained by any generous interpretation of little-C. Specifically, one can speak of the process of creativity and discovery and show that some aspects of it appear to require the postulation of an entity that is larger than an emergent mind. Particularly noteworthy are the extraordinary numerical coincidences in the history of science that are probabilistically next to impossible [45].

We add that the consideration of information (or entropy) in physical theory, which is commonly done in many branches of physics, implies an unstated postulation of consciousness. Information cannot be reduced to local operations by any reductionist program. It requires the use of signs derived from global properties and the capacity to make choices which, in turn, implies agency. Such agency will be consistent with physical law only if does not involve the expenditure of energy.

### 5. MEASUREMENT, COLLAPSE, DECOHERENCE

Let us consider the interaction problem in the framework of quantum theory. The measurement operation divides the physical universe into two parts: the first part is the system being observed, and the second part is the human observing agent, together with the measurement apparatus.

Philosophically, space and time must divide for the universe to be born. Likewise, the experimenter must be separated from the system being observed. In the orthodox Copenhagen Interpretation (CI) [26][27], a hypothetical interface called the Heisenberg cut (or the von Neumann cut) is assumed between quantum events and the observer's information. Below the cut everything is governed by the quantum wave function, whereas above the cut one must use classical description. The cut might appear to be arbitrary, but it is merely a way to separate parts of the system, which must be done in a consistent manner.

The CI considers the question of interaction between mental states and the wave function by taking the wave function to have an epistemological reality, that is, it represents the experimenter's knowledge of the system, and upon observation there is a change in this knowledge. Operationally, it is a dualist position, where there is a fundamental split between observers and objects. The placement of the cut between the subject and the object is arbitrary to the extent it depends on the nature of the interaction between the two.

In the ontic view of the wave function, there is no collapse of the wave function, and the interaction is seen through the lens of decoherence, which occurs when states interact with the environment producing entanglement. Decoherence causes the system to make transition from a pure state to a mixture of states that the observer is able to measure [46]. The process of decoherence in no way negates the CI picture, for it merely shifts the cut away in such a way that the system under observation and the measurement apparatus are on the same side.

Let us consider the system to be in the state $|\varphi\rangle = \sum_i c_i |\varphi_i\rangle$ with two components (so that $i = 0$ or 1). Let the measurement apparatus be described by $|\rho\rangle = 1/\sqrt{2} \sum_i |\rho_i\rangle$, $i = 0,1$. The two interact and their joint state function is $|\varphi\rangle|\rho\rangle$.

The composite system evolves and gets entangled so that we have the mixed state with diagonal terms in the density matrix that is $|c_0|^2 |\varphi_0\rangle|\rho_0\rangle + |c_1|^2 |\varphi_1\rangle|\rho_1\rangle$. The measurement of the apparatus state reveals the state of the system although the

knowledge of the probability amplitudes is possible only by doing tomography on many identical copies of the original state.

By placing the Heisenberg cut away from both the system and the measurement apparatus, the problem of collapse of the wave function is sidestepped but it still involve the agency of the observer. It raises other questions: Since the entire universe may be taken to be a quantum system, the question of how this whole system splits into independent subsystems arises. It would seem that the splitting into subsystems, which is an observational choice, serves about the same function as the Heisenberg cut of CI. From the perspective of the mind, the ontic view is troublesome for its own states are determined by transformational operations that rule out agency, and to assume a mechanistic behavior for the mind is to oppose the inconvertible fact of free will.

Bohr argued that the consideration of the biological counterpart to the observation of the relation between mind and body does not become part of an infinite regress. He argued that [47] "We have no possibility through physical observation of finding out what in brain processes corresponds to conscious experience. An analogy to this is the information we can obtain concerning the structure of cells and the effects this structure has on the way organic life displays itself.... What is complementary is not the idea of a mind and a body but *that* part of the contents of the mind which deals with the ideas of physics and the organisms and *that* situation where we bring in the thought about the observing subject."

The experimenter is not describing reality ontologically; rather, he is obtaining knowledge about it and this depends on the nature of his interaction with the system. The knowledge informs his mind and consideration of this information creates a sense of overarching knowledge. If information is fundamental then the violation of Bell's theorem by experiments does not imply a fundamental difficulty [48].

### 6. QUANTUM ZENO EFFECT

If the physical world and consciousness did not interact then evolution of the universe will be chaotic. While the psychological part of the psychophysical parallelism notion implies that consciousness does not have a physical basis (it has physical correlates, which is not the same thing), the quantum Zeno effect [28] provides a mechanism on how observation can influence dynamics sidestepping the question of the ontological position of the observer.

This effect arises from the fact that a quantum system's evolution may be stopped by measuring it frequently enough with respect to some chosen measurement basis. In other words, a watched system does not change. Opposite to this, if a system is measured frequently enough along specific bases, its evolution can be guided to a desired state. The name of this effect is a take-off on Zeno's arrow paradox, according to which an arrow in flight is not seen to move during any single instant, and therefore it cannot possibly be moving at all.

The quantum Zeno effect does not change the dynamics, and the process of observation merely changes the probabilities that are associated with different outcomes. It finds a way for consciousness to manifest itself in evolution without the need for any change to the physical law.

The idea of samavāya (inherence) in the Vaiśeṣika Sūtra of Kaṇāda [31] is a similar idea in which consciousness influences the physical world by observation alone. It is extraordinary that this subtle idea has been a part of the mainstream philosophy in India for a long time.

### 7. IMPLICATIONS FOR MACHINES

Cognitive architectures for the solution of certain AI problems are organized around three computational areas of subconcepts (preconscious or subliminal), abstract concepts, and a linguistic interface (Figure 7) which is similar to the hierarchical agents model described by neuroscientists [20].

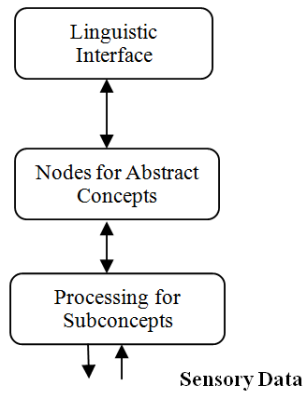

Figure 7. A cognitive architecture

The implementation of the engineered system must unpack the various subconcepts that are needed in the solution. The subconcepts could be various feature detection operators that feed into the higher-level operators as in a computer vision problem. For a robot, the issue is to have an inner map of the world together with the robot, so that it can navigate through the many obstacles that might be present. There is no need to have awareness within the system for that is already programmed in the design of the system (Figure 8).

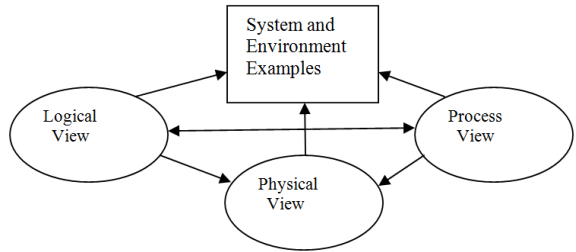

Figure 8. A cognitive architecture

This above view is consistent with general belief in the AI community that machines can replicate all cognitive tasks performed by conscious agents so long as these tasks are not based on the capacity of awareness, which is fundamental for self-reflection and creativity. But even if subjective awareness is not engineered, machines will replace humans in a huge variety of tasks, causing unprecedented stress and dislocations in society and raise questions of meaning and purpose of life.

Machines will come up short in creativity tasks, although the term "creativity" itself is contested and in retrospect what was taken to be creative at one point may be seen as a consequence of previous causes. Indeed, some creativity is inductive but if there is another kind which is non-inductive, then this will not be open to machines.

These ideas have implication to society for one may conclude that cultures that regiment human thought reduce its members to be no better than automatic, machine-like behavior. Such cultures diminish humanity and so they will come in conflict with open societies. On the other hand, the alienation set off in society due to the vanishing of jobs may attract some to cults with a simplistic view of the world.

## 8. DISCUSSION

We have shown that the naïve view of considering consciousness to be apart from the body provides surprising insight into the larger problem of conscious machines. Cognitive capacities are computational but their assignment to the autobiographical self is a process that is associated with awareness and memories. This assignment occurs with consciousness as a singular phenomenon. Sentience is a complex dance between being and becoming, where being is consciousness and becoming is the physical reality.

Consciousness cannot intervene in physical law but it can change the probabilities in the evolution of quantum processes (as in the quantum Zeno effect), without changing the dynamics and this provides an explanation of how consciousness can be reconciled with the physical law.

Let me return to the question of why the brain-machine is conscious. If the phenomenon of consciousness is contingent on a recursive and self-organizing structure that constitutes the unity of the organism, then we know that current machines will come up short. We don't yet know whether machines can be designed that will have such a structure for we lack a mathematical theory of computation for adaptive, self-organizing components. Perhaps a case could be made that only biological machines can have such a basis and that opens up the possibility of engineering new biological structures that have consciousness.